\documentclass[12pt]{article}
\usepackage[dvips]{graphics}

\begin{document}

\hspace{2cm}  

\medskip
\begin{center}

\Large{\bfseries 
Theory of the Anomalous Magnetic Moment of the Electron}

\vspace{0.5cm}

\large{E.L. Koschmieder}

\bigskip

\small{Center for Statistical Mechanics\\The University of Texas at 
Austin, Austin TX
78712.  USA\\
e-mail: koschmieder@utexas.edu}

\end{center}

\bigskip

\noindent

\small{It is shown that it follows from our model of the electron     
that its magnetic moment has an anomalous part, if the magnetic
field energy is taken into account. That means that the magnetic 
moment of our model of the electron is 1.000\,0565 times larger 
than the measured magnetic moment of the electron with its anomalous 
part.} 
\normalsize 

\section*{Introduction}

\bigskip
 
As is well-known Kusch and Foley [1] discovered that the magnetic 
moment of the electron is not equal to Bohr's magneton but differs 
from $\mu_B$ by an additional small amount on the order of one  
part in a thousand. This additional part is referred to as the anomalous 
magnetic moment. Soon afterwards Schwinger [2] proposed the first 
approximation of the theoretical explanation of the anomalous magnetic 
moment of the electron. This was followed by a long series of increasingly 
accurate measurements of the value of the anomalous magnetic moment, 
accompanied by increasingly accurate theoretical explanations of the 
anomalous magnetic moment. At present the magnetic moment of the 
electron is known with exceptional accuracy. It is
\begin{equation} \mu_e = 1.001\,159\,652\,1859 \pm 38\cdot 10^{-12}\, 
\mu_B\,,
\end{equation}
according to the Review of Particle Physics [3]. $\mu_\mathrm{B}$ is 
Bohr's magneton,

\bigskip
\hspace{4cm} $\mu_\mathrm{B}$ = e$\mathrm{\hbar}$/2mc.

\bigskip
\noindent 
Eq.(1) can, according to Mac\,Gregor [4], be approximated by
\begin{equation}\mu_e \cong \mathrm{\frac{e\hbar}{2m(e)c}}(1 + \alpha_f/2\pi) = 
\mathrm{\frac{e\hbar}{2m(e)c}}(1 + 0.001\,161\,409) = 1.000\,0018\,\mu_e(exp)\,,
\end{equation}
\noindent
with the fine structure constant $\alpha_f$. 

The QED explanation of $\mu_e$, which uses a bare part of the electron
with an infinite mass and an infinite negative charge, both of which are very 
questionable, practically matches 
the experimental results. One must wonder, on the other hand, 
how it is possible that the magnetic moment of the electron can be 
explained with such accuracy, when we do not possess an accepted 
explanation for neither the mass of the electron, nor for the spin  
of the electron, without which there would be no magnetic moment.

We have previously proposed an explanation of the structure of the electron
in [5], according to which the electron consists to one half of electron neutrinos 
$\nu_e$, complying with the age-old observation
of Poincar\'{e} [6] that the electron cannot consist exclusively of charge, 
because the charge elements  repel each other, which would render 
the electron unstable. The other half of the electron consists, in our model, 
of electric oscillations which carry the charge. Since neutrinos do not 
contribute to the magnetic moment of a particle the mass
 m(e) in the equation for the magnetic moment of the electron 
\begin{equation} \vec{\mu}_e = \mathrm{\frac{g\,e\hbar}{2m(e)c} \,\vec{s}}
\end{equation}
has to be changed. The ratio e/m, which classically describes the spatial 
distribution of charge and mass, has to be corrected. Eq.(3) contains the 
invariable constants e, $\mathrm\hbar$ and c. The mass m(e) of 
the electron is a constant too.
However, only the current, not the mass of a current loop, determines 
the magnetic moment of a loop. Therefore we can, for the calculation
of the magnetic moment of the electron, take into account only the part of
the mass of the electron which represents the charge, in order to arrive at 
a correct determination of the magnetic moment. The charged part of our
model of the electron is equal to m(e)/2. Since the spin of the 
electron is s = 1/2 it follows, if m(e) in Eq.(3) is replaced by 
m(e)/2, that the Land\'{e} factor g in Eq.(3) must be equal to one 
and that 
\begin{equation}\mu_e = \mathrm{e\hbar/2m(e)c} = \mu_B\,,
 \end{equation}
as it must be. To quote from p.63 of [5]: ``if exactly one half of the mass of 
the electron consists of neutrinos, then it follows automatically that the electron 
has the magnetic moment $\mu_e$ = e$\mathrm{\hbar}$/2m(e)c." But,  
as we will see, the charged part of the electron is not exactly 
equal to m(e)/2.

\begin{center}
\section{The magnetic moment of the electron} 
\end{center} 

The magnetic moment of the electron depends on the mass of the electron, as
expressed by Eq.(4). An accurate determination of the magnetic moment of 
the electron depends therefore on an accurate explanation of its mass. According
to the model of the electron we have proposed in Section\,11 of [5] one half   
of the mass of the electron consists of electron neutrinos, the other half of 
m(e) consists of electric oscillations which carry the charge e. The sum
of both parts does not leave room for another contribution to m(e). 
However, there is energy in the magnetic field of the 
magnetic moment, and this energy has to be taken into
account in the sum of the energies which make up the electron.

   The first calculation of the magnetic field energy of the electron was made by
Rasetti and Fermi [7].  They used the classical equations for the magnetic field 
caused by a current loop and obtained an equation for the field energy 
E$_\mathrm{H}$
\begin{equation} \mathrm{E_H} =\mu^2/3\mathrm{R}^3\,, \end{equation}
with the magnetic moment $\mu$ and the radius R of a sphere. The magnetic
field outside of R was integrated from R to infinity. Born and Schr\"{o}dinger [8]
suggested similarly that E$_\mathrm{H}$ = $\mu^2$/2R$^3$. Mac\,Gregor [4] 
finally suggested that the total magnetic field energy inside and outside of R is
 equal to \begin{equation} \mathrm{E_H} = 2\mu^2/ 3\mathrm{R}^3\,.
 \end{equation}
In the absence of a better equation we will use Eq.(6). Using for R the 
Compton wavelength R$_C$  = $\mathrm{\hbar}$/m(e)c  =
 3.8611$\cdot$10$^{-11}$ cm Mac\,Gregor arrived at a formula for  
the total magnetic field energy of the electron
\begin{equation} \mathrm{E_H} = \mathrm{\alpha_f/6\cdot m(e)c^2}\,.
\end{equation}
However, in our model of the electron the mass in the 
charged part of the electron is m(e)/2 so, instead of Eq.(7),  the magnetic 
field energy of the electron is actually
\begin{equation} \mathrm{E_H(e)} = \alpha_f/6\cdot\mathrm{m(e)c^2/2}
 = 0.00\,121\,622\cdot\mathrm{m(e)c^2}/2\,. \end{equation}

   In our model of the electron the sum of the masses of the electron neutrinos 
in the electron is equal to m(e)/2 and is fixed, and equal to the product of the 
constant number N/4 of the neutrinos in the lattice times their constant  
mass m($\nu_e$). The other half of the mass of the electron consists of the 
charged part of the electron m(e)$_{cp}$ plus the total magnetic field energy 
E$_\mathrm{H}$(e) divided by c$^2$. According to Eq.(8) roughly one part
in a thousand
of the energy in the electric oscillations is used for the magnetic field energy.
The sum of the electric charges remains equal to the elementary electric charge.
That means that the charged part of the electron m(e)$_{cp}$ must be 
\begin{equation} \mathrm{m(e)}_{cp} = \frac{\mathrm{m(e)}}{2
(1 + \alpha_f/6)} \cong \frac{\mathrm{m(e)}}{2}\cdot(1 - \alpha_f/6) \,.
\end{equation}
The energy in the charged part of the electron plus its magnetic field energy 
(Eq.7) is then m(e)c$^2$/2, as it should be, neglecting higher order terms of 
$\alpha_f$.

   The magnetic moment of the electron caused by 
the charged part of the electron is then, with Eq.(9),
\begin{equation} \mu_e = \frac{\mathrm{e\hbar}}{2\mathrm{m_{cp}(e)c}}\cdot
\mathrm{s} = \frac{\mathrm{e\hbar}\cdot2(1 + \alpha_f/6)}{2\mathrm{m(e)c}}
\cdot\mathrm{s} \,, \end{equation} 
and with s = 1/2 and $\mu_\mathrm{B}$ = e$\mathrm{\hbar}$/2m(e)c follows 
that \begin{equation} \mu_e(theor) = \mu_\mathrm{B}\cdot(1 + \alpha_f/6) =         
 \mu_\mathrm{B} \cdot1.001\,21622 \,, \end{equation}
whereas the actual value of the magnetic moment is $\mu_e$ = 
$\mu_\mathrm{B}$\,$\cdot$\,1.001\,15965, Eq.(1). 
The ratio of our theoretical value of the magnetic moment of the electron in 
Eq.(11) to the experimental value of the magnetic moment of the electron in 
Eq.(1) is then
\begin{equation} \mu_e(theor) = 1.000\,0565\cdot\mu_e(exp) \,. \end{equation}
A more precise determination of the magnetic moment depends on a more 
precise formula for the magnetic field energy.

   The sole reason for the deviation of the magnetic moment of the electron from
the classical value $\mu_\mathrm{B}$ = e$\mathrm{\hbar}$/2m(e)c is the 
incorporation of the magnetic field energy into our equations. Without the 
magnetic field energy the magnetic moment of the electron would 
be exactly equal to Bohr's magneton, as we have shown in Section\,12 of [5]. 

   Using for the explanation of the electron a relativistically spinning sphere,
which does not seem to be possible, Mac\,Gregor writes on p.\,315 of [4] that
``E$_\mathrm{H}$ is singled out as the culprit that is causing the anomaly in [the
Land\'{e} factor] g". It does not seem to be possible either that the electron 
consists of a bare part with an infinite mass and an infinite negative charge, as
QED assumes. We use a model of the electron whose mass is, within an 
uncertainty of one percent,  
equal to the measured rest mass of the electron. Nothing is infinite in
our model. There is indeed a bare part of the electron, but its mass is m(e)/2, 
rather than infinite. 

   The same considerations made above apply as well for the
magnetic moment of the positron, whose mass consists to one half of 
anti-electron neutrinos and to the other half of the mass of a positive 
charge and its magnetic field energy. 

\bigskip
 
\section*{Conclusions}

   We have shown that it is a straightforward consequence of our model
of the electron that the electron has an anomalous magnetic moment, if
the magnetic field energy is taken into account. Our model of the
electron assumes that one half of the mass of the electron consists
of electron neutrinos. The other half of the mass of the electron consists 
of the mass in the electric charge plus the magnetic field energy divided
by c$^2$. It turns out that in our model of the electron its magnetic moment
differs from the measured magnetic moment of the electron with its anomalous 
part by the factor 1.000\,0565, if we take the magnetic field energy into account.

\section*{References}

\noindent
[1] Kusch, P. and Foley, H.M. 1947. Phys.Rev. {\bfseries72},1256.

\smallskip
\noindent
[2] Schwinger, J. 1949. Phys.Rev. {\bfseries76},790.

\smallskip
\noindent
[3] Review of Particle Physics, 2011.

\smallskip
\noindent
[4] Mac\,Gregor, M.H. 2007. \emph{The Power of $\alpha$}. World Scientific.

\smallskip
\noindent
[5] Koschmieder, E.L. 2008. http://arXiv.org/physics/0804.4848.

\smallskip
\noindent
[6] Poincar\'{e}, H. 1905. Compt.Rend. {\bfseries140},1504.

\smallskip
\noindent
[7] Rasetti, F. and Fermi,E. 1926. Nuovo Cim. {\bfseries3},226.

\smallskip
\noindent
[8] Born, M. and Schr\"{o}dinger, E. 1935. Nature {\bfseries135},342.

\end{document}